\documentclass[11pt,english,fleqn]{article}

\usepackage[utf8]{inputenc}
\usepackage[a4paper, total={16cm, 23cm}]{geometry}
\usepackage{graphicx}
\usepackage{amsfonts}
\usepackage{amssymb}
\usepackage{amsmath}
\usepackage{epsfig}
\usepackage{amsthm}
\usepackage{babel}
\usepackage{multicol}
\usepackage{fancyhdr}
\usepackage{slashed}
\usepackage{eucal}
\usepackage{mathrsfs}
\usepackage[dvipsnames]{xcolor}
\usepackage{comment}
\usepackage{empheq}

\usepackage{cite}
\usepackage{authblk}

\graphicspath{{home/}}

\numberwithin{equation}{section}

\DeclareSymbolFont{bbold}{U}{bbold}{m}{n}
\DeclareSymbolFontAlphabet{\mathbbold}{bbold}

\makeatletter
\DeclareFontEncoding{LS2}{}{\noaccents@}
\makeatother
\DeclareFontSubstitution{LS2}{stix}{m}{n}
\DeclareSymbolFont{xlargesymbols}{LS2}{stixex}{m}{n}
\DeclareMathSymbol{\sumop}{\mathop}{xlargesymbols}{"B3}

\newcommand{\bo}{\boldsymbol}
\newcommand{\smallfrac}[2]{{\textstyle\frac{#1}{#2}}}
\newcommand{\BE}{\begin{equation}}
\newcommand{\EE}{\end{equation}}
\newcommand{\mrm}{\mathrm}

\newcommand{\dd}{\mathrm{d}}

\newcommand{\me}{\mathrm{e}}
\newcommand{\mcal}{\mathcal}
\newcommand{\mn}{\mathnormal}
\newcommand{\del}{\partial}

\newcommand{\mSig}{\mn\Sigma}

\newcommand{\intprod}{\boldsymbol{\mathbin{\raisebox{\depth}{\scalebox{1}[-1]{$\lnot$}}}}}

\makeatletter
\newsavebox{\@brx}
\newcommand{\llangle}[1][]{\savebox{\@brx}{\(\m@th{#1\langle}\)}%
  \mathopen{\copy\@brx\kern-0.5\wd\@brx\usebox{\@brx}}}
\newcommand{\rrangle}[1][]{\savebox{\@brx}{\(\m@th{#1\rangle}\)}%
  \mathclose{\copy\@brx\kern-0.5\wd\@brx\usebox{\@brx}}}
\makeatother

\title{\textbf{Quantization: History and Problems}}
\author{Andrea Carosso\thanks{acarosso@gwu.edu}}
\affil{The George Washington University\vspace{-2ex}}
\date{\today\vspace{-5ex}}

\begin{document}

\sloppy

\maketitle

\abstract{In this work, I explore the concept of quantization as a mapping from classical phase space functions to quantum operators. I discuss the early history of this notion of quantization with emphasis on the works of Schr\"odinger and Dirac, and how quantization fit into their overall understanding of quantum theory in the 1920's. Dirac, in particular, proposed a quantization map which should satisfy certain properties, including the property that quantum commutators should be related to classical Poisson brackets in a particular way. However, in 1946, Groenewold proved that Dirac's mapping was inconsistent, making the problem of defining a rigorous quantization map more elusive than originally expected. This result, known as the Groenewold-Van Hove theorem, is not often discussed in physics texts, but here I will give an account of the theorem and what it means for potential ``corrections" to Dirac's scheme. Other proposals for quantization have arisen over the years, the first major one being that of Weyl in 1927, which was later developed by many, including Groenewold, and which has since become known as Weyl Quantization in the mathematical literature. Another, known as Geometric Quantization, formulates quantization in differential-geometric terms by appealing to the character of classical phase spaces as symplectic manifolds; this approach began with the work of Souriau, Kostant, and Kirillov in the 1960's. I will describe these proposals for quantization and comment on their relation to Dirac's original program.  Along the way, the problem of operator ordering and of quantizing in curvilinear coordinates will be described, since these are natural questions that immediately present themselves when thinking about quantization.}

\tableofcontents


\section{Introduction}

In physics the word ``quantization'' can mean several related things. Most broadly, it can refer to any derivation of quantum mechanics, no matter how heuristic, and from whatever fundamental postulates. A more narrow meaning it can have is the program of obtaining quantum theory from classical theory, in particular. It sometimes also refers to the discrete nature of certain objects in quantum theory, like energies, compared with their continuous classical analogues. Perhaps most often, the word refers to the particular procedure of \emph{replacing} classical objects, like functions $f(q,p)$ on classical phase space, with quantum \emph{operators} $\hat f$, which play a major role in the mathematical formalism of quantum mechanics; this is the meaning we will use throughout this work. The operators obtained by quantization are used to define the time evolution of quantum states via the Hamiltonian operator, $\hat H$, and they also constitute  ``observables'' whose expectation values may be calculated in any given state. What's more, quantization is sometimes applied not just to classical \emph{functions}, but to the classical dynamical \emph{laws} themselves, thereby obtaining quantum dynamical laws. Thus quantization is a sort of \emph{mapping} of classical theories to quantum theories, and quantization establishes a correspondence between classical theories and quantum theories (though not necessarily a one-to-one correspondence).

The notion of a quantization map was first indicated by Dirac 96 years ago, in his first paper on quantum mechanics in November 1925, although one could argue the notion was implicit also in Heisenberg's preceding work. Soon after their work, Schr\"odinger published his first paper on wave mechanics, in which he proposed the time-independent wave equation for hydrogen, but he did not explicitly acknowledge the notion of quantization until March, in his paper which demonstrated the equivalence between his wave mechanics and the matrix mechanics of Heisenberg. All of this will be discussed at length in sections 2 and 3. Now, many modern introductory texts on quantum mechanics introduce quantum theories following a mixture of Dirac and Schr\"odinger's approaches, by regarding them as particular quantizations of classical theories. The paradigmatic example is the quantization of a classical theory with Hamiltonian
\BE
H(q,p) = \frac{p^2}{2m} + V(q),
\EE
where $q, \; p$ are Cartesian coordinates. One maps the coordinates to operators via
\BE
q \mapsto \hat q = q, \qquad p \mapsto \hat p = -i\hbar \frac{\del}{\del q}
\EE
(an input from Schr\"odinger's work), and notes that these operators satisfy the so-called \emph{canonical commutators}
\BE
[\hat q, \hat p ] = i \hbar \mathbb I,
\EE
which one regards as a quantum counterpart to the classical Poisson bracket $\{q,p\} = 1$ (an input from Dirac's work), and which constitutes a sort of \emph{justification} of the mapping above. The replacement of a classical quantity, like $q$ or $p$, by an operator, may be characterized as a mapping $\mathbb Q$ of functions $f$, so that $\mathbb Q_f = \hat f$. The mapping of the phase space coordinates is taken to imply that the Hamiltonian operator should be
\BE \label{HamOp}
\hat H = \frac{\hat p^2}{2m} + V(\hat q) = - \frac{\hbar^2}{2m} \frac{\del^2}{\del q^2} + V(q),
\EE
which determines the time evolution of the wave function according to the time-dependent Schr\"odinger equation,
\BE
i \hbar \del_t \psi = \hat H \psi.
\EE
As is well-known, though, Schr\"odinger himself did not (at first) arrive at his equation by invoking any notion of quantization like that described above, but rather by appealing, for inspiration, to classical Hamilton-Jacobi theory and its analogy to ray optics, as I will discuss in section 2. And Dirac, prior to the summer of 1926, had not internalized Schr\"odinger's notion of the quantum objects as differential operators, having instead focused on the structural relation between classical and quantum quantities and laws, as I will discuss in section 3. In itself, the procedure outlined above reproduces only the  mathematical formalism of QM, and does not also give one the probability interpretation of $\psi$ that is necessary to connect it to empirical phenomena.\footnote{One must also add the notion of spin in order to obtain complete non-relativistic descriptions of electrons and protons interacting electromagnetically. The introduction of spin is often thought of as an extra step beyond the conceptual input of quantization, being referred to as an intrinsically quantum phenomenon with no classical counterpart; Dirac, for example, suggested this idea in his book \cite{DiracBook}. We will see in section 6 that in fact spin systems can be thought of as quantizations of a particular kind of classical theory.} We will come back to this point shortly.


The concept of quantization is again applied when extending quantum mechanics to the relativistic domain, where one obtains the Klein-Gordon equation, although in order to describe particles with higher spin than zero, one must also introduce the notion of wave functions in non-trivial representations of the Lorentz group, which generalizes the notion of spin; doing so, one obtains the Dirac equation and its generalizations. And lastly, the philosophy of quantization is invoked when motivating the step to quantum field theory (QFT), where one quantizes classical \emph{field} theories according to the general conceptual scheme of quantization. Quantization is particularly important in this domain, as it has been found that many successful QFT's (apart from quantum electrodynamics) are quantizations of classical theories that have no classical empirical counterpart, making their starting point rather more abstract than that of non-relativistic QM. It is clear that the notion of quantization therefore plays a significant role in the writing-down of quantum theories. Although this role is often considered to be somewhat heuristic, the success of its usage in the simplest cases has led to considerable work being done over the years to define the quantization map \emph{rigorously}, some of which will be described later.\footnote{The search for such a rigorous mapping is sometimes called the \emph{problem of quantization}.}

It is legitimate to doubt whether the notion of quantization as a rigorous mathematical mapping is well-founded, however: one could argue that no such mapping should be expected to exist to begin with. It is certainly possible that whatever the arguments that historically led us to the Schr\"odinger equation (and its generalizations) were, the resulting theory, being ``more fundamental'' than classical mechanics, need have no relation to classical mechanics \emph{above and beyond} the relation of reducing to it in some limit, i.e. a mapping QM $\to$ CM. If a quantization map existed, then it would constitute a further, \emph{forward} relation CM $\to$ QM. This would be a rather odd and abstract feature of the world, being a mapping from an approximate description of the world to a more fundamental, microscopic description. This oddness would be lessened if the mapping CM $\to$ QM always implied that the classical theory one starts with is obtained as the classical limit of the quantum theory, but whether this is true depends on how one defines the classical limit of a quantum theory; it is true for example in the WKB approximation, so long as one starts with an appropriately simple wave function.\footnote{See chapter 6 of \cite{Holland} for a discussion.} If one instead denies the existence of a quantization map, the procedure of quantization must then be viewed as a merely practical or heuristic method for obtaining quantum theories; but the success of the procedure would have to be ultimately fortuitous (unless one has an alternative way to derive quantum theory, which implies the results of quantization by its own means).

One might contrast the concept of quantization with another principle, which plays a part in the generalization of special relativity (SR) to general relativity (GR), as well as the method for obtaining the simplest forms of gauge theories in field theory. Namely, in going from SR to GR, one can obtain the geodesic equation by replacing partial derivatives $\del_\mu$ in the special-relativistic Newton's Second law with covariant derivatives $\nabla_\mu$. Likewise, in obtaining a minimally-coupled gauge theory from a theory of matter alone, one replaces $\del_\mu$ by $D_\mu$ involving the gauge potential. Should quantization be regarded as a rule on par with these? On the face of it, it seems like replacing $q,p$ by their operator counterparts is a similar type of rule. But note that in the context of GR or gauge theory, it is essential to work in terms of covariant derivatives: it is a reflection of the fundamental principles used to define the theory. In the case of quantization, the precise fundamental principle involved in justifying the replacement is not so clear. Dirac may be taken as offering such a fundamental principle, embodied in his suggestion of the commutators playing the role of a Poisson bracket; this will be discussed in section 3.

The problem of how rigorous the quantization map should be is distinct from the problem of interpreting the quantum formalism or the measurement problem. The standard mathematical formalism of quantum mechanics, involving certain differential operators acting on wave functions and the notion of expectation values, time-dependence, et cetera, is the goal of quantization, and these features do not obviously bear any insight into how the wave function is to be interpreted, or how measurements are supposed to be accounted for mathematically, or what  ontology should be ascribed to the world according to quantum theory. Although the introduction of the wave function seems to be a step beyond merely replacing classical observables by operators, one should note that the presence of operators implicitly assumes vectors to act upon; in this sense quantization does imply the presence of wave functions or quantum states in quantum theory. But again, their interpretation is not specified by the procedure. The philosophical problem that  quantization \emph{does} bring to the fore is the problem of how exactly classical mechanics and quantum mechanics are related, as mentioned previously --- of whether there exists a rigorous forward relation between them in addition to a backward relation. If there does exist a quantization map, then applying the map to obtain quantum theory might be counted as a \emph{derivation} of QM, or at least a derivation of the ingredients entering the mathematical theory (including, perhaps, the dynamical laws of QM), and may be compared with other attempts to derive quantum theory.

In what follows, the early history of the concept of quantization will first be reviewed. We will go over Schr\"odinger's path to wave mechanics, and how he began to use the notion of quantization as a tool for generating wave equations. I then discuss Dirac's philosophy of quantization and how he thought of the relation between CM and QM, and highlight some of the tensions in his approach. The Groenewold-Van Hove theorem will then be described, in section 4, which is commonly regarded as demonstrating the non-viability of Dirac's particular proposal for quantization. In sections 5 and 6 I relate in some detail a few modern attempts to define quantization in a rigorous fashion, namely, Weyl Quantization and Geometric Quantization, and comment on their relation to Dirac's original proposal. The notion of quantization, if accepted, naturally leads to questions about how curvilinear coordinates on the classical side are to be dealt with, as well as the problem of operator ordering; this will be discussed in section 7. Lastly, in section 8, the recent program of Geometric Quantum Mechanics will be mentioned, in which the algebraic aspect of Dirac's proposal takes the center stage, rather than the particular procedure of quantization. I will end the paper with some comments on the program of quantization.

\section{Schr\"odinger's Appeal to Quantization}

In his 1924 Ph.D. thesis, Louis de Broglie had postulated the existence of a \emph{matter wave} associated with every matter particle, inspired by Albert Einstein's earlier (1905) proposal that there may be photons, or particles of light, associated with classical electromagnetic waves. Erwin Schr\"odinger sought to determine how de Broglie's matter waves could be defined in general, and what their dynamics might be. In the first of his four initial papers on wave mechanics in January 1926, entitled ``Quantisation as a Problem of Proper Values, Part I,''\footnote{He used the word ``quantisation'' either as a general indication of obtaining quantum theory from a classical theory, or perhaps more specifically in the sense of obtaining integer values for certain quantities, like energy, but not in the sense of mapping particular classical objects into quantum objects.} he proposed a variational principle leading to what we now call the time-independent Schr\"odinger equation \cite{SchroCollected}.\footnote{See \cite{Joas2009} for an account of how the optical-mechanical analogy \emph{did} play a role in Schr\"odinger's first communication.} In the second paper, one month later, he admitted that his earlier argument had been ``unintelligible,'' and proposed a more principled derivation of his equation, based on the similarity of classical Hamilton-Jacobi (HJ) theory to ray optics (a similarity that Hamilton himself was well-aware of in formulating his mechanics back in the 1830's \cite{Joas2009}). We shall briefly review Schr\"odinger's argument below, and then we will discuss how the concept of quantization emerged in his work in 1926.\footnote{See \cite{Joas2009,Wessels} for more thorough historical accounts of Schr\"odinger's path to wave mechanics, including the major role that Einstein's gas theory played in its development. And see \cite{Mandelstam} or Schr\"odinger's original papers \cite{SchroCollected} for a fuller mathematical account of the derivation we present here.}

The central dynamical object in Hamilton-Jacobi theory for a point particle on configuration space $Q$ is Hamilton's principle function, $S(q,t)$, which is the classical action $S[q_c(t')]$ evaluated along a solution $q_c(t')$ of the Euler-Lagrange (EL) equations of motion, with final position $q$ at time $t$.
Rather than solving the EL equations to determine $q_c(t)$, one may instead solve the Hamilton-Jacobi partial differential equation
\BE \label{HJE}
\del_t S + H(q, \del_q S, t) = 0
\EE
for $S(q,t)$ and use the fact that the initial and final particle momentum in HJ theory are given by
\BE
p_0 = - \frac{\del S}{\del q_0}, \quad p = \frac{\del S}{\del q}
\EE
to solve for the trajectory $q_c(t)$. Now, for time-independent Hamiltonians, the HJ equation may be separated by writing $S(q,t) = W(q) - Et$. Schr\"odinger postulated that for particles of definite energy $E$, the phase of the matter wave $\psi$ should be simply $S(q,t)/\hbar$ where $\hbar$ is Planck's constant. Thus the wave fronts of the matter wave are surfaces of constant $S$, and these fronts propagate with a phase speed determined from the condition $\delta S = \delta W - E \delta t = 0$, and given by
\BE
v(q) = \frac{E}{\| p \|} = \frac{E}{\sqrt{2m(E-V(q))}},
\EE
where we have used the fact that the normal vector field to the wave front is $p(q,t) = \del_q W(q)$, and assuming a Hamiltonian $H = p^2/2m + V(q)$; thus the HJ equation for this case reads $E = (\del_q W)^2/2m + V(q)$. Schr\"odinger then assumed the wave $\psi$ satisfies the \emph{classical} wave equation with nonhomogeneous phase speed $v(q)$,\footnote{Schr\"odinger worked in a configuration space of generalized coordinates with arbitrary metric, but I use Cartesian coordinates throughout this section for simplicity.}
\BE
\del_t^2 \psi(q,t) = v(q)^2 \del^2_q \psi(q,t).
\EE
Plugging in the expression for the phase speed and assuming all time-dependence in $\psi$ comes from the phase yields (what we call) the time-independent Schr\"odinger equation,
\BE \label{SE}
E \psi(q) = - \frac{\hbar^2}{2m} \del^2_q \psi(q) + V(q) \psi(q),
\EE
where $\hbar$ is a dimensionful constant necessitated by requiring a dimensionless phase $S/\hbar$ for the wave. The leap to this equation was not rigorous, mathematically, since solutions to this equation do not generally possess $W(q)$ as the position-dependent part of their phase.\footnote{Under suitable conditions, the phase may nonetheless be equal to $W(q)$ \emph{to leading order} in $\hbar$; this is formalized by the WKB approximation.} This is analogous to the fact that wave optics cannot be obtained rigorously from ray optics; the latter is a limit of the former, namely, the limit of small wavelength, or negligible diffraction. 

In the fourth paper of the series, submitted in June of 1926, Schr\"odinger generalized this equation to time-dependent cases by looking for the simplest wave-like equation that is independent of $E$, finally settling on the equation
\BE \label{TDSE}
i \hbar \del_t \psi(q,t) = - \frac{\hbar^2}{2m} \del^2_q \psi(q,t) + V(q) \psi(q,t).
\EE
It is clear that the leap to a time-dependent equation for $\psi(q,t)$ is a significant one, representing a shift away from the optical-mechanical analogy which had led him to the time-independent equation, since the waves $\psi(q,t)$ satisfying the time-dependent Schr\"odinger equation do not need to simultaneously satisfy the classical wave equation. Note furthermore that from the point of view of the optical-mechanical analogy, as it is invoked in the second communication, the presence of the Laplacian in the Schr\"odinger equation has nothing to do with the particle momentum, in contrast with the point of view of quantization.

The notion of replacing $p$ by $- i \hbar \del_q$ did not appear in Schr\"odinger's first two papers on wave mechanics in 1926. It appears in his published work for the first time in his March 1926 paper on the equivalence between Heisenberg's matrix mechanics and his own wave mechanics \cite{SchroCollected}, which came out before parts III and IV of his ``Proper Values'' papers. In remarking on Heisenberg's use of functions of both position $q$ and momentum $p$, while in wave mechanics only functions of $q$ appear, he states
\begin{quote}
``So the \emph{co-ordination} has to occur in such a manner that each $p_l$ in the \emph{function} is to be replaced by the operator $\del/\del q_l$.'' (pg. 47 of \cite{SchroCollected})
\end{quote}
He states that such a replacement can be done for any function written as a power series in $q,p$, and goes on to show that one can obtain matrices from any such operator by taking appropriate inner products with respect to any complete set of functions of $q$. Later in the paper, notably, he appeals to his variational derivation of the wave equation rather than the optical-mechanical analogy, before remarking that it is consistent with what one obtains upon replacing $p$ by $-i\hbar \del_q$. This is because in the variational derivation there \emph{is} a more direct relation between the $p$ in the Hamiltonian and the appearance of $-i\hbar \del_q$ in the wave equation: One sets $S = k \log \psi$ in the HJ equation (with $k$ a constant), and since $p = \del_q S = k \del_q \psi / \psi$, there is a derivative of $\psi$ associated with each $p$. Setting the HJ equation (multiplied by $\psi^2$) as the integrand of a variational principle and minimizing, one finds the Schr\"odinger equation, and the two derivatives in the Laplacian term correspond to the two factors of $\del_q \psi$ in the Hamiltonian. I remark that if $H$ contains higher powers of momentum, like $p^3$ for example, the variational procedure leads to a \emph{nonlinear} Schr\"odinger equation, whereas textbook quantum mechanics would suggest the time evolution should still be linear. In this respect, Schr\"odinger's variational procedure disagrees with traditional quantization for Hamiltonians more than quadratic in momenta.

He invokes the notion of quantization again in his fourth paper on wave mechanics, mentioned above, when he derives the relativistic wave equation in a background electromagnetic field, which we now call the Klein-Gordon (KG) equation. He states:
\begin{quote}
``From the [relativistic Hamilton-Jacobi equation in an electromagnetic field] I am now attempting to derive the \emph{wave equation} for the electron, by the following \emph{purely formal} procedure, which, we can verify easily, will lead to [the non-relativistic wave equation], if it is applied to the Hamiltonian equation of a particle moving in an arbitrary field of force in ordinary (non-relativistic) mechanics.'' (pg. 119 of \cite{SchroCollected})
\end{quote}
He is saying that the ``formal procedure'' of replacing $p_\mu = \del_{q^\mu} S \to -i\hbar \del_{q^\mu}$ in the HJ equation, where $(q^\mu)=(t,q)$, yields both the relativistic KG equation \emph{and} the non-relativistic Schr\"odinger equation. He appears to have  resorted to this formal prescription because the optical-mechanical analogy fails to yield a suitably simple wave equation \cite{Joas2009}, requiring consideration of waves in a flowing medium.\footnote{Note, however, that one need not invoke quantization to obtain the \emph{free} KG equation: The optical-mechanical analogy in this case yields the desired equation. Indeed, Schr\"odinger's unpublished first derivation of the KG equation was along these lines, building off of de Broglie's work \cite{Kragh1984}.} Thus at this point the concept of quantization was establishing itself in his thoughts as a general method for obtaining quantum-mechanical wave equations.

It is interesting that he resorted to the quantization map so quickly after proposing the optical-mechanical analogy that led to the time-independent equation. We have seen above that there are two main reasons for this, coming from the desire to relate wave mechanics to Heisenberg's matrix mechanics, and the derivation of the relativistic wave equation. In the second paper, he explicitly states that the postulation that $\psi$ satisfies the classical wave equation is merely one of simplicity:
\begin{quote}
``The only datum for [the wave equation's] construction is the \emph{wave velocity} [\dots] and by this datum the wave equation is evidently not uniquely defined. It is not even decided that it must be definitely of the second order. Only striving for simplicity leads us to try this to begin with.'' (pg. 27 of \cite{SchroCollected})
\end{quote}
Thus, Schr\"odinger's later appeal to the quantization map may perhaps be reconciled with the optical-mechanical analogy simply by viewing it as a mechanism for generating wave-like equations, together with the fact that the optical-mechanical analogy fails to determine a unique wave equation for any given classical system, whereas using the classical Hamiltonian to obtain the wave equation does yield a unique operator (at least if we ignore the issue of operator ordering for Hamiltonians involving products of $q$ and $p$).

Lastly, we can ask what sort of mathematical properties the quantization mapping, as suggested by Schr\"odinger in his fourth paper, would possess. The general prescription would be to start with the classical HJ equation, and perform the mapping
\BE
q^\mu \mapsto \hat q^\mu = q^\mu, \quad p_\mu = \frac{\del S}{\del q^\mu} \mapsto \hat p_\mu = - i \hbar \frac{\del}{\del q^\mu}.
\EE
In the standard case with $H = p^2/2m + V(q)$, performing this mapping in the HJ equation (\ref{HJE}) yields the operator equation
\BE
i\hbar \frac{\del}{\del t} = - \frac{\hbar^2}{2m} \frac{\del^2}{\del q^2} + V(q),
\EE
which is assumed to generate time evolution of the wave function upon application to $\psi(q,t)$. In effect, then, such a procedure yields a quantization mapping that satisfies two properties,
\begin{align}
\mrm{Linearity \; rule}: & \quad \mathbb Q_{af+bg} = a \mathbb Q_f + b \mathbb Q_g \nonumber \\
\mrm{Power \; rule}: & \quad \mathbb Q_{f^n} = (\mathbb Q_f)^n
\end{align}
where $a,b$ are constants on phase space. He assumed these two properties implicitly in his paper about matrix mechanics, when he remarked that replacement of $p$ by $-i\hbar \del_q$ can be done for functions written as power series in $p$. It is not clear, however, whether Schr\"odinger believed in the notion of quantization as a \emph{rigorous} mathematical mapping that relates classical and quantum theories, or whether it was merely a heuristic rule for obtaining wave equations, as mentioned above. If his later work in the 1930's on providing a statistical basis\footnote{See \cite{Zambrini:1985} for a review and development of Schr\"odinger's ideas in this direction.} for the Schr\"odinger equation is any indication, however, it is likely that for him the procedure of quantization, as described above, was more heuristic than not.

For Paul Dirac, on the other hand, the notion of quantization played a central role in his understanding of quantum theory, as we will now explore.



\section{Dirac's Program of Quantization}

In July of 1925, six months before Schr\"odinger's first paper on wave mechanics, Heisenberg submitted his famous paper on quantum mechanics, which introduced the notion of Matrix Mechanics. It became apparent in the subsequent work of Born and Jordan in September that in the quantum theory ``non-commuting'' variables played an essential role, wherein position and momentum were to be regarded as matrices (e.g. $\bo q = [q_{nm}]$ for position).\footnote{I use boldface letters to denote matrices in the next few equations.} Moreover, the position and momentum matrices were found to satisfy the ``canonical commutation relation'' (CCR),
\BE
[\bo q, \bo p] = i \hbar \bo 1,
\EE
where $\bo 1$ is the identity matrix. Furthermore, the time evolution of these matrices was determined by what are now called the Heisenberg equations of motion,
\BE
\frac{\dd \bo q}{\dd t} = \frac{1}{i \hbar} [\bo q, \bo H], \qquad \frac{\dd \bo p}{\dd t} = \frac{1}{i \hbar} [\bo p, \bo H].
\EE
The necessity of introducing matrices into quantum theory and thinking of classical quantities like position and momentum as non-commuting objects was mysterious to some physicists \cite{Pais}. On November 7 of 1925, however, Dirac's first paper on QM was received, ``The Fundamental Equations of Quantum Mechanics,'' in which he proposed a general conceptual scheme for understanding quantum theories consisting of the two related notions of (1) algebraic equivalence and of (2) quantization of classical quantities. These ideas were further elucidated in his book, \emph{The Principles of Quantum Mechanics}, first published in 1930 \cite{DiracBook}. I draw on both the 1925 paper and his book in what follows, while stressing where his views differ between the two.

Dirac tried to make sense of the newly-discovered QM by supposing that the fundamental relationship between classical and quantum theory is that the quantum theory should \emph{preserve the algebraic structure} of classical mechanics (CM), while relaxing the commutativity of the dynamical variables of CM. I refer to this as his \emph{algebraic philosophy} towards QM; he referred to it in his book as the ``method of classical analogy.'' Now, the object which determines the algebraic structure of CM is the Poisson Bracket (cPB), which acting on arbitrary functions $f,g$ on phase space, is given by
\BE
\{ f, g \} = \frac{\del f}{\del q} \frac{\del g}{\del p} - \frac{\del f}{\del p} \frac{\del g}{\del q},
\EE
and this object satisfies various algebraic identities: antisymmetry $\{f, g\} = - \{g, f\}$, linearity in both arguments, a Leibniz rule $\{f,gh\} = g\{f,h\} + \{f,g\}h$, and the Jacobi identity
\BE
\{ f, \{ g, h \} \} + \{ g, \{ h, f \} \} + \{ h, \{ f, g \} \} = 0.
\EE
Thus Dirac sought to find a ``quantum'' Poisson Bracket (qPB) corresponding to the cPB, by assuming that it satisfies all the same algebraic laws as the cPB, except that it operates on non-commuting objects and therefore no variables may be commuted through each other in manipulating the identities above. He argued that the most general such object was given by
\BE \label{qPB}
\mrm{qPB}(X,Y) = \frac{1}{i \hbar} \big( XY - YX \big),
\EE
where $X$ and $Y$ are two non-commuting objects in the quantum theory, and $\hbar$ is some constant with units of action, to be determined empirically. I will denote the commutator $XY-YX$ by $[X,Y]$ in what follows. At this stage, there's no necessary relationship between the quantum objects $X$ and classical objects such as position or momentum, but if there \emph{is} an algebraic relationship between the structure of QM and CM, a natural assumption to make is that there is a correspondence between the objects which obey the two algebras. If $x$ is a classical function on phase space, its quantum counterpart may then be denoted $X = \hat x$ (or $\mathbb Q_x$), and $\hat x$ may be referred to as the \emph{quantization} of $x$.\footnote{The notion of $\hat x$ or $\hat p$ as \emph{operators} was not explicitly introduced until Schr\"odinger's matrix-wave equivalence paper, a few months after Dirac's paper was received.} In deciding on the particular relationship between a quantum object $\hat x$ and its classical counterpart, he proposed (in his book) that
\begin{quote}
``The strong analogy between the quantum P.B. [\dots] and the classical P.B. [\dots] leads us to make the assumption that the quantum P.B.'s, or at any rate the simpler ones of them, have the same values as the corresponding classical P.B.'s.'' (pg. 87 of \cite{DiracBook})
\end{quote}
I will call this the ``Dirac rule'' in what follows. In terms of equations it would read literally as
\BE \label{qPBeqcPB}
\frac{1}{i\hbar} [\hat x, \hat y] = \{x,y\}.
\EE
To say that the qPB has ``the same value'' as the cPB is ambiguous, however; what he meant was for the right-hand side to be the \emph{quantum} object corresponding to the classical function $\{x,y\}$, namely, $\mathbb Q_{\{x,y\}}$. Thus, since classical position and momentum satisfy $\{q,p\}=1$, the quantum counterparts should satisfy\footnote{Along with $[\hat q, \hat q]/i\hbar = 0, \; [\hat p, \hat p]/i\hbar = 0$, since $\{q,q\}=\{p,p\}=0$ classically.}
\BE \label{CCR}
\frac{1}{i\hbar} [\hat q, \hat p] = \hat 1.
\EE
The assumption that quantum theory preserves all the algebraic laws of the classical theory, and that there is a correspondence between classical quantities and quantum quantities, therefore implies the canonical commutators of Heisenberg, Born, and Jordan. Moreover, since Hamilton's equations may be written in terms of the cPB as
\BE
\frac{\dd q}{\dd t} = \{q, H\}, \quad \frac{\dd p}{\dd t} = \{p, H\},
\EE
application of the Dirac rule immediately yields the Heisenberg equations of motion (assuming that quantization and time derivatives commute). Indeed, in the 1925 paper he states\footnote{The remark about ordering in this quote is dispelled, later in the paper, by noting that products of $q$ and $p$ never occur for Hamiltonians of the $T(p) + V(q)$ form (pg. 75 of \cite{DiracCollected}). But this is a coordinate-dependent statement, and moreover, one might nonetheless be interested in observables that do involve such products, even if the Hamiltonian does not. Dirac considered the problem of quantization in polar coordinates in his second paper, as I discuss in section 7.}
\begin{quote}
``We are now able to take over each of the equations of motion of the system into the quantum theory provided we can decide on the correct order of quantities in each of the products. Any equation deducible from the equations of motion by algebraic processes not involving the interchange of the factors of a product [\dots] may also be taken over into the quantum theory.'' (pg. 70 of \cite{DiracCollected})
\end{quote}
Hence, the algebraic philosophy provided a basis for understanding the structure of the new quantum mechanics, which otherwise was quite unfamiliar to classical physics.

By assuming an association between a classical observable $x$ and a quantum object $X=\hat x$, Dirac invoked the notion of a quantization map. He noted that the commutators $[\hat f, \hat g]$ of operators corresponding to general functions $f, g$ of $q$ and $p$ could be evaluated by expressing $\hat f, \hat g$ in power series in $\hat q, \hat p$ and repeatedly using the CCR, thus yielding the ``quantum conditions'' for $\hat f, \hat g$. This implicitly assumes the linearity and power rules for quantization that we identified in the previous section on Schr\"odinger's quantization scheme. Moreover, he assumed that the classical function $1$ maps to the identity,
\BE
\mrm{Identity \; rule}: \quad \mathbb Q_1 = \mathbb I
\EE
(this is also implied by Schr\"odinger's quantization method, if the classical Hamiltonian contains a constant term). Now, in the case of $q$ and $p$, classically $\{q,p\}=1$, so there's no ambiguity in arriving at the CCR, $[\hat q, \hat p] = i \hbar \mathbb I$. If $\{f,g\}$ is a nontrivial function of $q,p$, however, then presumably the mapping of its power series to a quantum operator should appear on the right-hand side of the commutator of $\hat f$ with $\hat g$. It is not obvious that this operator is the same as that obtained by his repeated-CCR algorithm mentioned above. It is also clear that just the linearity rule and power-rule are not sufficient to determine the quantum operator $\hat f$ in general, because there's no rule for specifying the order of products like $q^n p^m$. In general, different orderings will differ by operators proportional to $\hbar$ (due to the CCR); and thus the commutator rule would generally hold only to ``leading order'' in $\hbar$.\footnote{I use quotes here because even if two operators only differ by powers of $\hbar$, the (fractional) difference between their expectation values need not be small, depending on the states they are evaluated in. This is discussed in section 7 below.}
These subtleties are probably the reason he added the note of caution that eq. (\ref{qPBeqcPB}) holds for at least the ``simpler'' quantities, rather than all quantities. We will come back to the problem of operator ordering in section 7.

In his book, he urges that the method of classical analogy is not applicable to all quantum systems (pg. 84 of \cite{DiracBook}). He may be referring to the fact that some systems, like spin systems, seem to have no classical phase space dynamics which they may be regarded as the quantization of (but this is not necessarily true, as will be discussed in section 6). After describing how the CCR's may be used to obtain the quantum conditions for any two functions $f,g$, he states that the CCR's ``thus give the solution of the problem of finding the quantum conditions, for all those dynamical systems which have a classical analogue and which are describable in terms of canonical coordinates and momenta. This does not include all possible systems in quantum mechanics'' (pg. 88 of \cite{DiracBook}). Thus he seems to have viewed the CCR's as sufficient conditions for quantizing any classical phase space; for cases like spin systems, however, the method of classical analogy is not expected to be applicable.\footnote{For example, on pg. 143 of \cite{DiracBook} he writes, ``The spin does not correspond very closely to anything in classical mechanics, so the method of classical analogy is not suitable for studying it.''}



In many texts on quantum mechanics it is stated that the classical theory is recovered in the limit $\hbar \to 0$, because in this limit the canonical commutators become $[\hat q, \hat p] = 0$, signifying that $\hat q, \; \hat p$ then commute, as they do classically. In Dirac's book he states this definitively:
\begin{quote}
``Equations [eq. (\ref{qPB})] and [eq. (\ref{CCR})] provide the foundation for the analogy between quantum mechanics and classical mechanics. They show that \emph{classical mechanics may be regarded as the limiting case of quantum mechanics when $\hbar$ tends to zero}.'' (pg. 88 of \cite{DiracBook})
\end{quote}
If one takes Dirac's proposal seriously that the quantum commutator plays the role of a Poisson bracket, then there would seem to be a tension in regarding this as the definition of the classical limit: for position and momentum the limiting case of the commutators is $[\hat q, \hat p] = 0$ while classically $\{q, \; p\}=1$, so that the $\hbar \to 0$ limit seems to eliminate the Poisson bracket from the theory altogether. In other words, the commutators seem to play the dual role of the quantum Poisson bracket \emph{and} measuring the deviation from classical commutativity, i.e. the quantum conditions. This tension is slightly resolved by making sure to distinguish the $\mrm{qPB}$ from the commutator: since $\mrm{qPB}(\hat q, \hat p) = \mathbb I$ regardless of the value of $\hbar$, it may then hold as $\hbar \to 0$, while forcing $[\hat q,\hat p]\to 0$ at the same time. But it is not clear if the qPB \emph{reduces} to the cPB in this limit, however, which one would expect for consistency; in his November 1925 paper, Dirac had an argument for how this might happen in the context of matrix mechanics (indeed, this observation served as part of the motivation for his proposing the Dirac rule). Later in the paper, however, Dirac indicated that the $\hbar \to 0$ limit was not particularly significant. He writes,
\begin{quote}
``The correspondence between the quantum and classical theories lies not so much in the limiting agreement when $h \to 0$ as in the fact that the mathematical operations on the two theories obey in many cases the same laws.'' (pg. 74 of \cite{DiracCollected})
\end{quote}
This statement is much in line with the algebraic philosophy towards quantum theory, but we saw in the previous quote that he later appealed to the $\hbar \to 0$ limit of the commutators in his book. It may be that the later work on the WKB expansion led him to reconsider the importance of the $\hbar \to 0$ limit. Indeed, when he discusses the classical limit of QM in more detail in chapter 5 of his book, he appeals to the WKB method, in which $\hbar$ being relatively small plays an important role in the expansion.

In his November paper, Dirac was thinking of quantum quantities, following Heisenberg, as matrices of ordinary commuting numbers. In his second paper about QM (received January 22, 1926), he explicitly distinguishes between classical and quantum quantities by introducing the notion of ``c-numbers'' and ``q-numbers.'' At this point he seems to have been reluctant to specify the precise mathematical and physical nature of q-numbers; he says, ``At present one can form no picture of what a q-number is like'' (pg. 88 of \cite{DiracCollected}), though later he states that under certain circumstances, they may be represented as matrices of c-numbers. For him, the algebraic properties of the q-numbers (via non-commutativity and the qPB) were their fundamental properties. Indeed, he went on in that paper to give a solution to the quantum Kepler problem without ever specifying a realization of the q-numbers involved. The reason for associating certain q-numbers with classical quantities, like position or momentum, was that they entered the quantum equations in a way that is formally the same as in the classical equations, via the cPB's.

The notion of quantum objects as differential operators did not enter Dirac's writings until the summer of 1926. Dirac discusses Schr\"odinger's work in print for the first time in his sixth paper about QM, ``On the Theory of Quantum Mechanics," received in August 1926. In an earlier work\footnote{His fourth paper about QM, ``Relativity Quantum Mechanics with an Application to Compton Scattering.'' See pg. 137 of \cite{DiracCollected}.} he had described how time $t$ and its canonical conjugate $p_t = - H$ could be treated as quantum variables on par with position and momentum, satisfying their own CCR's. Now Dirac remarked that specifying the coordinates $q^\mu$ to be usual numbers but letting $p_\mu$ be differential operators, resulted in the Schr\"odinger wave equation when a classical equation like $p_t + H(q,p,t) =0$ was thought of as a constraint equation on functions $\psi$. Later in the paper (after postulating what we now call Fermi-Dirac statistics), he describes how the same reasoning leads, in the relativistic case, to the Klein-Gordon equation. In discussing Schr\"odinger's contribution to QM, Dirac mostly appealed to the variational principle (see pg. 179 of \cite{DiracCollected}), which he considered to be a special case of his general approach, for which the momentum variables become differential operators. For Dirac, presumably, any realization of the position and momentum operators would do, so long as they satisfied the CCR's: ``These relations form the main justification for the [assumption $p_\mu \mapsto -i\hbar \del_\mu$].'' (pg. 181 of \cite{DiracCollected}).\footnote{In his book (pg. 92 of \cite{DiracBook}), he explores a more general realization of the momentum operator, $\hat p = - i \hbar \del_q + f(q)$, concluding that the resulting quantum mechanics is gauge-equivalent to the Schr\"odinger representation. This exercise is highly suggestive of the later notion of prequantization, which we will meet in section 6.} The Stone-von Neumann theorem is the modern expression of this sentiment: the Schr\"odinger representation of the position and momentum operators is unique, up to unitary transformations \cite{Hall2013}. In taking the quantization of $p_t + H(q,p,t) = 0$ to be true in the quantum theory, Dirac seems to be supplementing his program of quantization, since the classical constraint equation is not an algebraic law of CM, in the sense of being formulated in terms of a cPB. To the extent that the Schr\"odinger equation is equivalent to the Heisenberg equations of motion, however, the constraint equation need not to be assumed; indeed, Dirac presents his own demonstration that the Schr\"odinger equation is equivalent to matrix mechanics.\footnote{Dirac later put much effort into incorporating constraints on the classical side into the quantization procedure \cite{DiracLectures}. This led to his constraint formalism for classical Hamiltonian mechanics, which involved modified Poisson brackets now known as \emph{Dirac brackets}. The constraints, together with the algebraic relations they imply involving the Dirac bracket, must be carried over into the quantum theory.}


As mentioned in the introduction, quantization is sometimes regarded as a way to \emph{derive} quantum mechanics from classical mechanics. A more precise (but not rigorous) statement of the argument, in light of Dirac's program of quantization, might run as follows. If we postulate the algebraic equivalence of classical and quantum laws while relaxing the commutativity of the variables constrained by those laws, one is led to look for an algebraic object providing the definition of the quantum PB's; commutators are the natural candidate. Since cPB's are classical algebraic laws, the qPB's must then satisfy the Dirac rule. The Dirac rule applied to position and momentum then yields the CCR. In itself, the algebraic equivalence principle does not specify what exactly the quantum objects are. A particularly simple realization of the quantum objects corresponding to $q,p$, however, are the usual Schr\"odinger operators $\hat q = q, \; \hat p = -i\hbar \del_q$. As differential operators they naturally act on functions of $q$, which may be called wave functions. Time evolution of the quantum operators is then given by the Heisenberg equations of motion, according to the algebraic equivalence. Equivalently, time evolution can be regarded as occurring for wave functions $\psi$ satisfying the constraint equation $\big(\hat p_t+H(\hat q, \hat p, \hat t)\big)\psi = 0$. If the quantization map relating classical observables to quantum operators satisfies the linearity, identity, and power rules, then one obtains the usual time-dependent Schr\"odinger equation.

The argument above hinges, in part, on the properties of the quantization map. Modern treatments of quantization tend to interpret Dirac as proposing that the quantization of  arbitrary classical observables should satisfy four rules,\footnote{One also sometimes adds the requirements (1) that real functions are mapped to self-adjoint operators, and (2) that $\hat q, \hat p$ are the usual Schr\"odinger operators, or that the operators act irreducibly on wave functions, meaning that their operation on position space wave functions again yields wave functions depending only on position, and not position \emph{and} momentum.}
\begin{align}
\mrm{Linearity \; rule}: & \quad \mathbb Q_{af+bg} = a \mathbb Q_f + b \mathbb Q_g \nonumber \\
\mrm{Power \; rule}: & \quad \mathbb Q_{f^n} = \hat f^n \nonumber \\
\mrm{Identity \; rule}: & \quad \mathbb Q_{1} = \mathbb I \nonumber \\
\mrm{Dirac \; rule}: & \quad [\mathbb Q_f, \mathbb Q_g] = i \hbar \mathbb Q_{\{f,g\}}
\end{align}
where $f,g$ are functions, and $a,b$ constants, on phase space. The last property makes $\mathbb Q$ into a homomorphism, mathematically, from the classical Lie algebra of observables (with bracket $\{\bullet, \bullet\}$) to the quantum algebra (with bracket $[\bullet, \bullet]$); this gives a literal meaning to the notion that QM should preserve the algebra of CM, and thus captures Dirac's algebraic philosophy. However, I emphasize again, it is not likely that Dirac actually thought the fourth rule should hold in general, but only for the ``simpler'' operators, as we saw before. As I remarked earlier, the link between the algebraic philosophy and quantization involves the additional assumption that there should be a correspondence between classical and quantum quantities. In section 8, I will review a proposal for understanding QM which relaxes this assumption.

One can nonetheless inquire about the internal consistency of the four rules above. And it turns out one can prove that no such $\mathbb Q$ map exists. This result is called the Groenewold-van Hove theorem.

\section{The Groenewold-Van Hove Theorem}

In October of 1946 Hilbrand Groenewold published his Ph.D. thesis ``\emph{On the Principles of Elementary Quantum Mechanics}'' \cite{Groenewold1946} whose intent was largely to support the impossibility of giving a description of quantum mechanics as the statistical mechanics of some underlying theory on phase space with ``uniquely determined processes,'' in the spirit initiated by von Neumann of producing no-hidden-variables theorems. Part of this demonstration, he argued, depended on there being a correspondence between classical observables and quantum observables. In this context he ultimately proved the nonexistence of any quantization map satisfying the properties Dirac suggested. Then later, in 1951, the physicist L\'eon Van Hove proved a more general version of Groenewold's result \cite{vanHove1951}, and in the modern literature their work is referred to as the Groenewold-Van Hove (GVH) theorem. Regardless of Groenewold's conclusions about the viability of hidden variable theories, his result \emph{does} undermine the Dirac quantization proposal of the previous section.

The GVH theorem can take several forms in the mathematical literature, but here we present the theorem (more or less) as Groenewold presented it, which is particularly simple (see [12] for another statement of the proof). The strategy is to deduce a contradiction from the four quantization rules listed in the previous section.\footnote{Groenewold does not assume the factorization rule as stated above, but instead assumes that if $a \to \hat a$, for $a$ a function on phase space, then $f(a) \to f(\hat a)$, which he attributes to von Neumann’s formulation of the postulates of QM. If $f$ has a power series expression, then linearity and the power rule imply this rule.} It is a simple exercise to show that a contradiction is produced for a particular function, $f(q,p) = q^2p^2$. We can determine the operator $\mathbb Q_{q^2p^2}$ using only the four quantization rules: First notice that we can write $f$ in terms of a (classical) Poisson bracket,
\BE
q^2 p^2 = \{q^3, p^3\}/9.
\EE
The Dirac rule (and linearity) then implies
\BE
\mathbb Q_{q^2 p^2} = \mathbb Q_{\{q^3, p^3\}/9} = \frac{1}{9 i \hbar} [\mathbb Q_{q^3}, \mathbb Q_{p^3}].
\EE
The right-hand side may be simplified using the properties of commutators, together with the power rule and the CCR (which follows from the Dirac rule), to obtain
\BE
\mathbb Q_{q^2 p^2} = \hat q^2 \hat p^2 - 2 i \hbar \hat q \hat p + \smallfrac{2}{3} (i\hbar)^2.
\EE
But notice that we could have also used
\BE
q^2 p^2 = \{q p^2, q^2 p\}/3
\EE
to decompose $f$; we can then write each function $q p^2, q^2 p$ in terms of even simpler cPB's, which (after further use of commutator identities) yields
\BE
\mathbb Q_{q^2 p^2} = \hat q^2 \hat p^2 - 2 i \hbar \hat q \hat p + \smallfrac{1}{3} (i\hbar)^2.
\EE
The two expressions of $\mathbb Q_{q^2p^2}$ differ at order $\hbar^2$, thus yielding a contradiction. It follows that at least one of the four rules of quantization above is not warranted. Indeed, the various combinations of assumptions that are mathematically consistent have been worked out \cite{Ali:2004}.\footnote{If the Dirac rule is denied, however, one must  add an assumption to the effect that $\hat q, \hat p$ are the usual Schr\"odinger operators $q,  -i\hbar\del/\del_q$.} It should be noted that if we restrict the space of functions to quantize to be the at-most quadratic functions in  $q,p$, then one \emph{does} obtain a quantization free of contradictions, in Dirac's sense; indeed, one obtains the quantum Heisenberg-Weyl algebra (see, e.g.\cite{Woit}). This algebra therefore tends to play an important role in the study of quantization.

In the 1925 paper Dirac argues that the consistency of the commutator rule is evidenced by the initially reasonable argument that, since the cPB's and qPB's obey the same algebraic laws, and the cPB's are consistent, the qPB's must then be consistent. By ``consistent'', in the case of the quantum theory, he means that it is impossible to deduce the vanishing of $\hbar$ from the commutator rule: ``The possibility [\dots] of deducing by quantum operations the inconsistency $h = 0$ thus cannot occur.'' (pg. 74 of \cite{DiracCollected}). But this is precisely what GVH deduces, if we regard $\hbar$ as an a priori unknown constant! As we saw in the previous section, Dirac's later account of quantization in his book was more cautious about the scope of the commutator rule; perhaps he was aware of problematic cases like $f=q^2p^2$ --- it is hard to imagine Dirac not having considered such operators. But he does not mention any such problematic operators in his early papers on quantum mechanics, except to say that subtleties arise when there is no unique ordering for the operators of interest, as noted in the previous section.

The inconsistency of the Dirac rule takes some of the steam out of Dirac's appeal to the algebraic equivalence of CM and QM. One cannot (rigorously) quantize the classical Hamilton's equations to obtain the Heisenberg equations, unless one adds the ad hoc caveat of restricting it to Hamiltonians of a sufficiently simple form.\footnote{Namely, the at-most quadratic functions in $q,p$. And even among such ``simpler'' Hamiltonians, the quantization mapping must be carried out in Cartesian coordinates on phase space in order to be without ambiguity. This will be discussed in section 7.} Proponents of the idea that quantization is nothing more than a heuristic tool might therefore take the GVH theorem as evidence for their view: since the natural-seeming rules one wants of a quantization map lead to inconsistencies, one might conclude that no rigorous map exists. A response to this conclusion is that the Dirac rule may be nothing more than a simplest suggestion, in which case different maps with different variations of the Dirac rule might yet be consistent (but perhaps coming at the cost of relaxing the role of algebraic equivalence). Indeed, numerous proposals to ``correct'' Dirac's program have arisen over the years, the most popular of which seems to be \emph{Weyl Quantization} (WQ), which proposes a specific ordering prescription for all polynomials in $q, p$, and finds much use in the literature. WQ satisfies a generalized Dirac rule where the right-hand side becomes a series in $\hbar$, with only the leading term producing $i \hbar \mathbb Q_{\{f,g\}}$. I give a brief account of WQ below, and in section 6 we will describe another ``modern'' quantization method. See \cite{Ali:2004} for a more thorough review of quantization methods in the mathematical literature.

\section{Weyl Quantization}

Weyl quantization was introduced by Hermann Weyl in 1927, and subsequently developed by John von Neumann and Eugene Wigner separately in the early 1930's (see \cite{Zachos} for details and references). In this case the quantization map $\mathbb Q^W_f$ is defined on polynomials first by assuming $\hat q, \hat p$ are the usual Schr\"odinger operators, and then by mapping $q^n p^m$ to a linear combination of all possible orderings of the $n \; \hat q$'s and $m \; \hat p$'s amongst each other with equal weights. So for example,
\BE
\mathbb Q^W_{q^2 p} = \frac{1}{3} \big( \hat q^2 \hat p + \hat q \hat p \hat q + \hat p \hat q^2 \big).
\EE
The Weyl operator $\mathbb Q^W_f$ for arbitrary $f$ can be equivalently characterized by \cite{Hall2013}
\BE
\mathbb Q^W_f = \frac{1}{(2\pi)^2} \int \dd a \dd b \; \tilde f(a,b) \; \exp\Big(ia \hat q + i b \hat p \Big),
\EE
where $\tilde f(a,b)$ is the Fourier transform of $f(q,p)$. This operator can be expressed in terms of an integral kernel acting on wave functions $\psi(q)$ if one works in the position space representation. Wigner's contribution was to discover the inverse of the Weyl map \cite{Zachos}, which, for a certain class of functions, determines a one-to-one correspondence between $f$'s and $\hat f$'s. Now, it can be demonstrated that for $f,g$ two functions on phase space, the composition of their Weyl operators, $\mathbb Q^W_f \mathbb Q^W_g$, is again a Weyl operator corresponding to a unique function $f \star g$, 
\BE
\mathbb Q^W_f \mathbb Q^W_g = \mathbb Q^W_{f \star g}.
\EE
The non-commutative product $f \star g$ is known as the \emph{Moyal product} of $f$ and $g$, and can be expressed in terms of the Fourier transforms of $f,g$. The Moyal product is often thought of as a \emph{deformation} of the classical product of functions, since one has
\BE
\lim_{\hbar \to 0} f \star g = fg.
\EE
An early form of the $\star$-product was initially discovered by von Neumann in 1931, but it was Groenewold and Moyal (separately) who first used it extensively in the late 1940's \cite{Zachos}.\footnote{I remark that the work of Groenewold and Moyal forms the basis for the \emph{phase space formulation} of quantum mechanics \cite{Moyal:1949,Zachos}.}

Weyl quantization does not satisfy the Dirac rule for all $f,g$, but it does satisfy it for $f$ a polynomial of degree at most 2 in $q,p$ and $g$ an arbitrary polynomial \cite{Hall2013}. Furthermore, there is a generalized version of the Dirac rule that it does satisfy exactly. It turns out that the antisymmetrization $\{\!\{f,g\}\!\} := f \star g  - g \star f$ of the Moyal product satisfies an identity,
\BE \label{moyalbracket}
 \{\!\{f,g\}\!\} = \sumop_{n=0}^\infty \frac{1}{n!} \Big(\frac{i \hbar}{2} \Big)^n \big( \Pi_n(f,g) - \Pi_n(g,f) \big),
\EE
where $\Pi_n$ is a differential operator known as the ``Poisson bivector'', which generalizes the Poisson bracket to include higher order derivatives, and is equal to the Poisson bracket for $n=1$. The left-hand side above is known as the \emph{Moyal bracket} (although Groenewold's paper appeared before Moyal's \cite{Zachos}). Mapping both sides by $\mathbb Q^W$ then yields a generalized Dirac rule where the right-hand side is a linear combination of Weyl operators, only the leading term of which is the quantization of $\{f,g\}$:
\BE
[\mathbb Q^W_f, \mathbb Q^W_g] = i \hbar \mathbb Q^W_{\{f,g\}} + O(\hbar^2),
\EE 
a relation we might call the \emph{Weyl rule}, after its resemblance to the Dirac rule. 

The Weyl map receives much attention in the mathematical literature due to its rigorous characterization and its relation to deformation theory. However, whether it constitutes a physically-preferable quantization map is not clear. For example, it does not quantize the squared-angular momentum correctly \cite{Zachos, Kleinert}; thus it is not necessarily the ``correct'' ordering prescription (I will discuss operator ordering in section 7).\footnote{In \cite{Kleinert}, a more pragmatic approach to quantization is proposed based on symmetries, to the effect that generators of the symmetry on the classical side are mapped to ``angular momentum'' differential operators on the quantum side. These operators constitute a representation of the Lie algebra associated with the symmetry, thereby satisfying the same Lie brackets. As such, this approach retains a restricted notion of algebraic equivalence.} It does have the virtue of being generalizable to arbitrary so-called \emph{Poisson manifolds}, which are smooth manifolds $M$ equipped with a Lie bracket (acting on the functions on $M$) which also satisfies the Leibniz rule; this theory was worked out by Kontsevich in 2003 \cite{Kontsevich}. In this generalized setting, Weyl quantization is referred to as \emph{Deformation Quantization}.

In a sense, Weyl quantization constitutes a shift away from Dirac's algebraic philosophy. Rather than preserving the standard algebraic structure of CM (the cPB), it uses an alternative, or deformed, algebraic structure associated with classical phase space, namely, the Moyal bracket, and bases the algebraic equivalence on \emph{that} structure. Thus, the relation between CM and QM under Weyl quantization is not as direct as the one envisaged by Dirac; one cannot say that CM and QM ``satisfy the same laws'' from the point of view of WQ, though they may be put in one-to-one correspondence.\footnote{In a work of Hiley and de Gosson \cite{Hiley-deGosson}, WQ is used to prove that there is a sense in which the Schr\"odinger equation (with Weyl-quantized Hamiltonian) can be derived from Hamiltonian mechanics. In particular, they prove that there is a one-to-one correspondence between Hamiltonian flows (see section 6) generated by a function $H(q,p)$ and solutions to the Schr\"odinger equation with Hamiltonian operator $\mathbb Q^W_H$.} Another drawback of WQ is that it takes the operators $\mathbb Q^W_q = q$ and $\mathbb Q^W_p = -i\hbar \del_q$ (or unitary equivalents thereof) as assumptions, rather than deriving them. The choice may be \emph{justified}, however, by noting that they satisfy the CCR, and that the CCR is implied by the Weyl rule when applied to position and momentum. It should be noted that demanding the operators $\hat q, \hat p$ satisfy the CCR does not quite determine them to be unitarily equivalent to the accepted Schr\"odinger operators; one must further specify that they act \emph{irreducibly} on the wave functions of the system. We will see an example of a realization of the position and momentum operators that do not act irreducibly in section 6.

A different proposal for quantization, called \emph{Geometric Quantization} (GQ), seeks to define a quantization map by appealing to differential-geometric and symplectic structures on classical phase space, and constitutes a more constructive approach to quantization than WQ.

\section{Geometric Quantization}

This approach began with the independent work of Kostant, Souriau, and Kirillov in the 1960's. They took as their motivation the problem of making quantization into a rigorous mathematical mapping. For example, of his work on the subject, Kostant later wrote
\begin{quote}
``In the early 60s I became interested in Hamiltonian mechanics and its symplectic manifold and Poisson bracket underlying structure. I also thought it was
quite mysterious and marvelous that physicists in quantizing classical mechanics
converted scalar functions (classical observables) on phase space in some fashion
or other to operators on Hilbert space. Particularly striking in this process was that
the classical observables were functions of position and momentum, $q$’s and $p$’s,
whereas the elements in the Hilbert space were “functions” on half the variables
(e.g., the $q$’s or the $p$’s). It seemed to me it would be very interesting to be able to
make this process rigorous.'' (pg. 533 of \cite{KostantCollected})
\end{quote}
Souriau expressed a similar sentiment, saying of Dirac's quantization proposal that
\begin{quote}
``Taken literally, these principles can have but a heuristic value; they are at the same
time contradictory (they must be weakened in order to obtain a coherent theory)
and incomplete (implicit complementary hypotheses arise when one applies them
in concrete cases). [\dots] The search for a theory mathematically coherent and physically achievable, which would constitute a “rational quantum mechanics,” is the program of geometric quantization.'' (pg. 600 of \cite{SouriauSelf})
\end{quote}
What resulted was an intricate mathematical formalism that was able to handle many of the most common cases of interest to physics, but was also sufficiently general to apply to a wider class of classical systems, in a sense to be described below. GQ differs in spirit from Weyl Quantization by giving a mathematical account of the construction of the Hilbert  space of a quantum system together with the operators acting on that space. Rather than postulating a particular form of $\hat q$ and $\hat p$ from the outset, as in Weyl Quantization, for example, these operators arise from a sequence of steps that starts with the classical phase space and the vector fields on that space. I now give a brief review of the formalism.\footnote{For more detailed introductions, see \cite{Todorov, Blau:2007, Nair:2016, Carosso:2018} or the textbooks \cite{Woodhouse, Abraham, Souriau1970}.}

Geometric quantization begins with the observation that there exists an operator for each function $f(q,p)$ on classical phase space, which \emph{does} satisfy the Dirac rule: namely, the operator
\BE
\mathbb P_f = i \hbar \frac{\del f}{\del q} \frac{\del}{\del p} - i \hbar \frac{\del f}{\del p} \frac{\del}{\del q} + f - p \frac{\del f}{\del p},
\EE
satisfies, for all differentiable $f$,
\BE
[\mathbb P_f, \mathbb P_g] = i \hbar \mathbb P_{\{f,g\}}.
\EE
But notice that since $\mathbb P_f$ is a first-order differential operator for all $f$, the mapping $\mathbb P$ does not satisfy the power rule of Dirac quantization. Thus in GQ one first constructs a pseudo-quantum theory based on this notion of quantization, before attempting to correct it to obtain a ``full'' quantization. The operator above is not unique, and it turns out that the operator $\mathbb P_f$ can be written in an intrinsic, coordinate-free way. But to understand this form of $\mathbb P_f$, as well as the rest of GQ, we must review some aspects of symplectic manifolds and Hamiltonian dynamics.

A symplectic manifold $M$ is a manifold of even dimension, $\dim M = 2n$, on which is defined a 2-form $\omega$ that is nondegenerate ($X \intprod \omega = 0 \Rightarrow X = 0$) and closed ($\dd \omega = 0$).\footnote{For lack of space, I must omit an introduction to the basic ingredients of differential geometry, such as vector fields, differential forms, and integration on manifolds. See \cite{Schutz} for a good introduction. I use the notation of differential geometry in order to emphasize the coordinate-independence of the formalism.} Darboux's theorem \cite{Arnold} implies that on any patch of $M$, we can find a ``canonical'' coordinate system $(q,p)$ such that
\BE
\omega = \dd p_a \wedge \dd q^a, \quad a = 1, \dots, n.
\EE
Classical phase spaces, being cotangent bundles $T^* Q$ for whatever given configuration space $Q$, constitute symplectic manifolds, where the canonical coordinate system covers the entire manifold. Now, by Poincar\'e's Lemma we can write $\omega = \dd \theta$, and the 1-form $\theta = p_a \dd q^a$ is called the symplectic potential. But other choices of $\theta$ are also valid, like $\theta = - q^a \dd p_a$. However, any two choices are related by a total differential, $\theta' = \theta + \dd u$, for $u$ a function.

Given a symplectic 2-form and a function $f$ on a symplectic manifold, the simplest way to associate a gradient-like vector field $X_f$ to $f$ is to set
\BE
X_f \intprod \omega + \dd f = 0,
\EE
since $\omega$ can be thought of as a map from vectors to 1-forms; $X_f$ is called the \emph{Hamiltonian vector field} of $f$. In canonical coordinates $X_f$ has the form
\BE
X_f = \frac{\del f}{\del p_a} \frac{\del}{\del q^a} - \frac{\del f}{\del q^a} \frac{\del}{\del p_a}.
\EE
The Poisson bracket can then be written variously as $\{f,g\} = X_g[f] = \omega(X_f, X_g)$. Now, the integral curves of $X_H$, if $H$ is some chosen Hamiltonian function, are then determined by Hamilton's equations:
\BE
\dot q^a = \frac{\del H}{\del p_a}, \quad \dot p_a  = - \frac{\del H}{\del q^a}.
\EE
Thus Hamiltonian mechanics can be thought of as the study of flows generated by Hamiltonian vector fields on a symplectic manifold.

Now back to quantization. Given some chosen $\theta$, the \emph{prequantization} of $f$ is defined by
\BE
\mathbb P_f = - i \hbar X_f - \theta(X_f) + f,
\EE
and this is a coordinate-independent object, which still satisfies the Dirac rule. It does, however, contain an arbitrary dependence on the \emph{gauge} chosen for $\theta$. That is, two different choices of $\theta$ will lead to different $\mathbb P_f$. However, it is a simple exercise to show that
\BE
\mathbb P_f^{(\theta_1)} \big[ \me^{-iu/\hbar} \psi \big] = \me^{-iu/\hbar} \mathbb P_f^{(\theta_2)} \big[ \psi \big],
\EE
when $\theta_2 = \theta_1 + \dd u$. This observation suggests that the wave functions we associate with this ``prequantum'' system should be taken as \emph{sections of a line bundle} over $M$, whose connection 1-form is $\theta / \hbar$, and therefore whose curvature is $\Omega = \omega / \hbar$.\footnote{The necessity of introducing Planck's constant in GQ therefore stems from the fact that the argument of the phase of the transition functions between different prequantization gauges must be dimensionless, while the symplectic 2-form $\omega$ is dimensionful. The value of $\hbar$, of course, is not determined by the formalism alone.} The prequantization of $f$ is then written in terms of a covariant derivative: $\mathbb P_f = - i \hbar \nabla_{X_f} + f$. In this way, the choice of potential $\theta$ amounts to a choice of gauge for the wave functions, in much the same way that ordinary wave functions in the presence of an electromagnetic field are sections of a bundle whose connection is the gauge potential $A = A_\mu \dd x^\mu$.\footnote{The presence of an EM potential is in fact incorporated quite naturally into the framework of GQ, where one starts from a ``charged'' symplectic 2-form $\Omega = \omega + e F$, $F$ being the magnetic field strength tensor.} Here, however, $\theta$ is non-dynamical, whereas $A$ is usually thought of as determined by Maxwell's equations. Thus wave functions in GQ, even absent the presence of an EM field, are sections of a bundle.

The natural Hilbert space of the prequantum theory is the space of square-integrable sections over $M$, with inner product
\BE
\langle \psi, \chi \rangle = \int_M \psi^* \chi \; \frac{\omega^n}{(2 \pi \hbar)^n},
\EE
where $\omega^n$, which is the standard Louiville measure on phase space, is just the $n$-th wedge power of $\omega$; the normalization of the measure allows one to work with dimensionless wave functions \cite{Woodhouse}. Now, not all symplectic manifolds can be prequantized, in the sense that in order for a line bundle with curvature $\omega/\hbar$ to exist over a symplectic manifold $M$, the 2-form must satisfy the so-called \emph{Weil integrality condition},
\BE
\int_\mSig \frac{\omega}{2 \pi \hbar} = n,
\EE
for all closed 2-surfaces $\mSig$ in $M$, where $n$ is an integer \cite{Woodhouse}. Cotangent bundle phase spaces trivially satisfy this condition with $n=0$. But for other symplectic manifolds, other values of $n$ are needed; for example, for $M = S^2$ the 2-sphere, $n$ ranges over all integers, depending on how many times $\mSig$ wraps around the 2-sphere.

The program of GQ is to modify the procedure of prequantization in a general way such that one obtains a full quantization, in the sense that the standard quantum mechanical theories are reproduced. For example, for a cotangent space $T^*Q$, one wants the quantization to yield the Schr\"odinger equation for any given classical Hamiltonian $H(q,p)$, and wave functions that live only on position or momentum space, rather than the full phase space. At this point the formalism quickly becomes highly technical, so I only provide a brief sketch of the construction:

\begin{enumerate}

\item We know from QM in practice that the wave functions are typically functions only of position $q$ or momentum $p$. A geometric way of saying this is that there is an $n$-dimensional set of vector fields $Y_a$ that is chosen to annihilate the wave functions: $Y_a[\psi] = 0$. Such a set of vector fields (subject to a few further restrictions) is called a \emph{polarization} of the manifold $M$. Thus for the position representation, $Y_a = \del/\del p_a$; this is called the \emph{vertical} polarization, since the $Y_a$ are along the fibers of $T^* Q$. But since $Y_a[\psi]=0$ is not gauge covariant, one must instead demand that the sections are covariantly annihilated, $\nabla_{Y_a} \psi = 0$, to eliminate the arbitrary dependence on the gauge. Thus ordinary position space QM corresponds under GQ to vertical polarization in the $\theta = p_a \dd q^a$ gauge.\footnote{Under GQ it is possible to have a position-space wave function (in the sense of being vertically-polarized) that nonetheless depends on momentum, if one uses a gauge in which $\theta(\del_p) \neq 0$.}

\item The prequantum inner product will generally diverge on polarized wave functions since there may not be any dependence on some of the coordinates being integrated over. To remedy this ailment one again redefines the wave functions to be not only polarized sections on $M$, but so-called polarized \emph{half-forms} on $M$, which may loosely be thought of as square-roots of volume-forms on half the space $M$. For example, for the vertically-polarized sections, the wave functions can be written as $\tilde \psi(q) = \psi(q) \sqrt{\dd q}$, were $\dd q$ is a volume-form on configuration space $Q$.

\item Even if one has a polarized half-form $\tilde \psi$, the half-form obtained by action of a prequantum operator on it, $\mathbb P_f \tilde \psi$, will generally \emph{not} be polarized. Thus one \emph{projects} the result back onto the polarization (using the notion of a so-called ``pairing'' on a symplectic manifold), a process known as Blattner-Kostant-Souriau (BKS) projection \cite{Woodhouse}. In the case of the vertical polarization of $T^* Q$, one obtains, for example, $\mathbb Q_{p^2} = - \hbar^2 \Delta + \smallfrac{\hbar^2}{6} R$, for $Q$ an arbitrary Riemannian manifold ($R$ being the Ricci scalar), and $p^2=g^{ij}p_ip_j$.\footnote{The presence of the Ricci scalar is a matter of debate, however. And of course, the term $\smallfrac{\hbar^2}{6} R \psi$ vanishes on flat configuration spaces, so the standard flat-space Schr\"odinger equation is produced. See \cite{DeWitt,Mostafazadeh} for more about this.} In this way GQ can reproduce the position space Schr\"odinger equation, eq. (\ref{TDSE}), by letting $\mathbb Q_H$ generate the time evolution of wave functions.

\end{enumerate}

There are several wrinkles in the program above, one of which is the failure to reproduce the constant ground state shift $\hbar \omega/2$ in the case of the harmonic oscillator. The resolution of this problem involves the introduction of the so-called ``metaplectic correction'' in GQ. Then there are more technical mathematical problems; for example, not all (pre)quantizable symplectic manifolds admit a polarization, and not all quantizations $\mathbb Q_f$  yield Hermitian operators. Nonetheless, the formalism of GQ is able to reproduce the most important case, namely, the Schr\"odinger equation for arbitrary potential $V(q)$ (and on arbitrary Riemannian configuration space). Quantum operators do not generally satisfy the Dirac rule in GQ (except for the subspace satisfying the Heisenberg-Weyl algebra), thereby avoiding the contradiction of the GVH theorem. However, one might not be very impressed by GQ if it must go to such great lengths just to obtain the usual Schr\"odinger-like quantum systems.

There is, however, a remarkable feature of GQ that deserves emphasis: it provides an algorithm for quantizing not just cotangent bundles, but many \emph{other} symplectic manifolds as well.\footnote{The quantization of classical \emph{field} theories falls under the $T^*Q$ category, since one is dealing with the cotangent bundle for $Q$ an infinite-dimensional manifold of field configurations. And of course, this still comes with all the mathematical subtleties of dealing with such spaces. See \cite{Woodhouse,Nair:2016} for examples.} For example, the 2-sphere $S^2$ is a symplectic manifold but not a cotangent bundle, and its quantization according to GQ yields the quantum mechanics of \emph{spin}! Specifically, the Weil integrality condition for $S^2$ yields a family of prequantizations $\mathbb P^{(n)}$ for $n \geq 0$ an integer. Choosing the so-called \emph{holomorphic polarization} to complete the quantization procedure in a given $n$-sector, one finds that the Hilbert space of square-integrable wave functions on $S^2$ is of dimension $n+1$. Now, the generator of rotations about the $z$-axis on the sphere is the Hamiltonian vector field of a function $J_3$ on $S^2$, and the quantization of this function yields an operator which, in the $n$-th prequantization sector, yields the standard spin-$z$ operator $\hat S_3$ with eigenvalues $-n,...,n$; thus one identifies $n = 2j$, where $j$ is the conventional spin quantum number. See \cite{Nair:2016,Carosso:2018} for more details on the quantization of spin in GQ.\footnote{A prominent perspective on GQ is offered via ``orbit theory,'' which thinks of GQ as a machinery for finding all the irreducible representations of a given Lie group (corresponding to the possible Hilbert spaces of the quantum system). The idea stems from the property that the \emph{coadjoint orbit} of any Lie group defines a symplectic manifold, which may then be quantized. For example, $S^2$ is the coadjoint orbit of $SU(2)$, and its quantization yields all the spin-$j$ irreps of $SU(2)$. See \cite{Woodhouse} for more on this.}

Thus, the quantization map as proposed by GQ is able to account for both quantizations of arbitrary $T^* Q$ (which includes the Schr\"odinger-type quantum systems on configuration spaces $Q$), as well as the phenomenon of quantum spin, the latter of which is typically regarded as an intrinsically quantum mechanical phenomenon. Dirac's statements about the inapplicability of the algebraic philosophy to spin systems were therefore somewhat hasty. Although spin is purely quantum mechanical in the sense that there is no \emph{empirical} classical counterpart to quantum spin, according to GQ there is nonetheless a classical \emph{theory} of spin, namely, Hamiltonian dynamics on the 2-sphere,\footnote{Similar conclusions have been drawn in other contexts as well, for example in the work of spin coherent states \cite{Radcliffe:1971} and their path integral formulation \cite{Altland}, or the work on classical models of the Dirac electron \cite{BarutZanghi}. I remark that spin in quantum field theory is usually accounted for by quantization of classical Grassmann-valued fields, but their spin structure is built-in from the start.} and the quantization of this system yields quantum spin systems.
One can even imagine an alternative historical scenario in which the phenomenon of spin was predicted a priori, after perhaps generalizing Schr\"odinger's quantization to other symplectic manifolds, in the manner of GQ above. Of course, this scenario is not very plausible, given the state of understanding of symplectic manifolds in the 1920's, and given the fact that the empirical trace of spin in atomic spectra was a glaring puzzle which was in any case likely to be solved in an ad hoc fashion, as indeed it was \cite{Pais}.

Regarding Dirac's algebraic philosophy of QM, GQ only sustains this idea on the level of prequantization, where it holds exactly. Thus in GQ, there is an algebraic equivalence, but it is between CM and prequantum mechanics, rather than full QM. In general the commutators of full quantum operators do not obey the Dirac rule, nor is there a systematic relationship between deviations from the Dirac rule and generalized Poisson brackets as there is in Weyl Quantization. We will see in section 8 that there is a sense in which the algebraic philosophy is true of QM, however, regardless of the quantization scheme used to obtain it (so long as it produces the standard mathematical formalism). But first, I will discuss the problem of dealing with curvilinear coordinates and operator ordering when quantizing, as well as the early history of these problems.

\section{Curvilinear Coordinates and Operator Ordering}

Classical Hamiltonian dynamics enjoys a large class of symmetries, namely, that the form of Hamilton's equations is invariant under certain transformations of the coordinates on phase space: the canonical transformations. Since canonical transformations preserve the cPB's, one could hope that Dirac's program will carry through just as simply in arbitrary canonical coordinates as for Cartesian $q,p$, since the CCR's in the new variables will still only involve the identity operator on the right-hand side. In particular, since any passive change of position space coordinates determines a canonical transformation, it may seem that quantization in arbitrary curvilinear coordinates should be achievable. It turns out, however, that attempting to do so leads to a host of problems, which have (typically) ad hoc solutions, as we will soon learn.

First I note that Dirac was aware of this difficulty. In a footnote in his book, for example, Dirac remarks that the replacement of $q,p$ in $H(q,p)$ by $\hat q, \hat p$
\begin{quote}
``[\dots] is found in practice to be successful only when applied with the dynamical coordinates and momenta referring to a Cartesian system of axes and not to more general curvilinear coordinates.'' (pg. 114 of \cite{DiracBook})
\end{quote}
In this sense, Dirac's program of quantization seems to \emph{care} about which coordinates we begin with on the classical side --- a situation many have regarded as unsatisfactory.\footnote{See \cite{Klauder:1997,Kalogeropoulos:2021} for perspectives on the importance of Cartesian coordinates in quantization.} Dirac came face to face with this problem earlier, back in his second paper about QM, received on January 22 of 1926, entitled ``Quantum Mechanics and a Preliminary Investigation of the Hydrogen Atom,'' in which he considered the Kepler problem. He began with the quantum Hamiltonian in terms of Cartesian coordinates, and then defined equations relating $(\hat x,\hat y,\hat p_x,\hat p_y)$ to $(\hat r,\hat \theta,\hat p_r,\hat p_\theta)$.\footnote{He used a slightly different notation from this.} Solving for $\hat p_x^2 + \hat p_y^2$ in terms of the polar quantities yielded the Hamiltonian
\BE
\hat H = \frac{1}{2m} \Big( \hat p_r^2 + \frac{1}{\hat r^2} \big( \hat p_\theta^2 - \smallfrac{\hbar^2}{4} \big) \Big) - \frac{e^2}{\hat r}.
\EE
But he noted that going to polar coordinates on the classical side and \emph{then} quantizing gives a different Hamiltonian,\footnote{Note that this is not an ordering issue, since $\hat r, \hat p_\theta$ in any case commute.}
\BE
\hat H = \frac{1}{2m} \Big( \hat p_r^2 + \frac{1}{\hat r^2} \hat p_\theta^2 \Big) - \frac{e^2}{\hat r}.
\EE
He remarks,
\begin{quote}
``The only way to decide which of these assumptions is correct is to work out the consequences of both and to see which agrees with experiment.'' (pg. 97 of \cite{DiracCollected})
\end{quote}
He ultimately chose the former expression, after finding that the latter expression leads to a doubling of all the energy levels of the system (and since this is ignoring spin, such doubling would not be observed experimentally).

We may consider the problem of quantizing in general curvilinear coordinates, where the metric $\mrm g_{ab}(q)$ is a function of $q$, following \cite{Kleinert}. Since a change of configuration coordinates is a canonical transformation on phase space, the cPB's are preserved, so that $\{q_a, p_b\} = \delta_{ab}$. Thus one would expect the new momenta to map to $\hat p_a = - i \hbar \del/\del q_a$, which preserves the CCR's. However, since the integration measure involves $\sqrt{\det \mrm{g}}$, the $\hat p_a$ as defined above will not generally be Hermitian.\footnote{The inner product is assumed to be $\langle \psi, \chi \rangle = \int_Q \dd q \; \sqrt{\det \mrm g} \; \psi^*(q) \chi(q)$.} Furthermore, assuming we want the quantization of $\mrm g^{ab}(q) p_a p_b$ to be (minus) the Laplacian on the configuration space, the naive mapping to $\mrm g^{ab}(\hat q) \hat p_a \hat p_b$ fails. One can redefine $\hat p_a$ to always be Hermitian and satisfy the CCR's, but one still does not obtain the Laplacian in general; one faces the problem of how to order the operator counterpart of $\mrm g^{ab}(q) p_a p_b$. Demanding Hermiticity is not sufficient to fix the operator uniquely, and many ``Hermitizations'' of the product do not yield the desired Laplacian, although there does exist an ordering which does so by inserting powers of $(\det \mrm g)^{\pm 1/4}$ around the $\hat p_a$ factors in a particular way. The resulting prescription is therefore a rather ad hoc procedure, and one has no guidance in situations where the Laplacian is not expected beforehand.\footnote{For general Riemannian spaces with curvature, there is the further problem of whether to include the Ricci scalar in the Schr\"odinger equation or not \cite{DeWitt}. It goes without saying that this issue is of some importance in the quantization of general-relativistic systems, as well as field theories with curved target spaces.} I remark that in practice, the tendency is to first quantize a problem in Cartesian coordinates, and then change coordinates on the level of the position-space Schr\"odinger equation, which then leads to the Laplacian in the new coordinates. We have seen in the previous section that in GQ, for example, one can obtain the correct Laplacian operator through the BKS projection method.

It is worth noting that Schr\"odinger's derivation of the wave equation has no problem with arbitrary curvilinear coordinate systems. Indeed, in his second communication, he worked in terms of such coordinates; the optical-mechanical analogy even works on Riemannian configuration spaces, since the wave fronts can be defined once one has a metric and therefore the notion of orthogonality and gradient vectors. Furthermore, in his paper on the equivalence of matrix and wave mechanics, he notes that the variational principle from his first communication can be easily generalized to arbitrary configuration spaces, which quite naturally then yields the time-independent Schr\"odinger equation involving the Laplacian in curvilinear coordinates as the Euler-Lagrange equation of the variational problem. See pg. 55 of \cite{SchroCollected} for his argument. Remember, however, that his variational approach does not reproduce the expected operators for higher powers than 2 in momenta.

Although the problem of quantization in curvilinear coordinates depends crucially on operator ordering, the issue of operator ordering in QM is a more general problem for quantization. That is, even for a theory quantized in Cartesian coordinates, say, one still faces an ordering problem for quantizing polynomials $p^n q^m$. As in the case of curvilinear coordinates, there are many Hermitizations of $\hat p^n \hat q^m$, so that demanding Hermiticity alone is not sufficient to fix the operator uniquely.\footnote{It is interesting to recall Dirac's reason for demanding that observables be Hermitian, as he describes in his book (pg. 35 of \cite{DiracBook}): If they were ``complex,'' then one generally could not measure both components simultaneously, due to Heisenberg uncertainty. But this assumes that the real and imaginary parts are canonically conjugate variables, which need not be true in general.} For example, for $f = p^2 q^2$ the operators
\BE
\hat f_1 = \frac{1}{2} \big( \hat p^2 \hat q^2 + \hat q^2\hat p^2 \big), \quad \hat f_2 = \hat p \hat q^2 \hat p,
\EE
are both Hermitian but not equal: $\hat f_1 = \hat f_2 - \hbar^2 \mathbb I$.\footnote{Different orderings of a given operator need not differ only by terms proportional to the identity, either. Take $\hat h_1 = (\hat q^3 \hat p^3+ \hat p^3 \hat q^3)/2$ and $\hat h_2 = (\hat q^2 \hat p \hat q \hat p^2 + \hat p^2 \hat q \hat p \hat q^2)/2$, which satisfy $\hat h_1 = \hat h_2 + 3 i \hbar^2 \hat q \hat p - \hbar^3 \mathbb I$.} The question of operator ordering quickly came up in the discovery of QM in 1925. In the paper of Born and Jordan in September \cite{SourcesQM}, they proposed a specific general ordering prescription, now called Born-Jordan quantization, which resembles that of Weyl quantization in the sense of proposing a certain ordering for polynomials, although they generally give different results for functions of order 2 or more in $q,p$. See \cite{deGosson} and references therein for more details.

Now presumably, given any particular physical system, the $q^2 p^2$ observable has some expectation value. Since the expectation value computed from the theory depends on the ordering one chooses, it would be natural that the question of operator ordering can then be solved experimentally, as Dirac had suggested in the case of quantizing in polar coordinates. This perspective has been proposed many times since the founding of quantum mechanics. We see it echoed in David Bohm's 1951 textbook, \emph{Quantum Theory}, when he states
\begin{quote}
``Until some experiment is found for which the predicted results depend on the method of Hermitization, there will be no way to decide which is the correct method.'' (pg. 186 of \cite{Bohm:1951})
\end{quote}
and later in Ramamurti Shankar's popular introduction to QM in 1980,
\begin{quote}
``There is no universal recipe for resolving such ambiguities [\dots] Symmetrization is the answer as long as [$f$] does not involve products of two or more powers of [$\hat x$] with two or more powers of [$\hat p$]. If it does, only experiment can decide the correct prescription.'' (pg. 120 of \cite{Shankar})
\end{quote}
One might expect that if the different possible operator orderings only differ at subleading order in $\hbar$, distinguishing between them experimentally will necessarily be difficult. I remark that even if two operators differ by a term of order $\hbar^2$, their expectation values in any given state need not differ only at subleading order. For example, if we consider the ground state of the 1-dimensional harmonic oscillator, then $\langle \hat f_2 \rangle = \frac{3}{4} \hbar^2$ , so $\langle \hat f_1 \rangle = -\frac{1}{4} \hbar^2$, and these are independent of the parameters $m, \omega$ in the Hamiltonian. Thus their expectation values can differ quite substantially, even though the operators differ by a term proportional to $\hbar^2$, and one can hope that experiment could therefore choose between them more easily than one would think. However, if the observable overall is order $\hbar^2$, of course, the signal may be difficult to detect to begin with.

We have not carried out a careful review of the literature to determine if there have been any direct experiments to decide on the issue of operator ordering, but here we comment on the implications of the problem of ordering. It seems possible that different physical systems may demand different ordering prescriptions, so that no one ordering accounts for all the instances of measurement of that observable; and if so, there would not seem be to any explanation for why one ordering was needed, but not another. And even for one and the same physical system, there may be multiple ways of measuring any particular observable, like $q^2 p^2$, each way yielding values corresponding to the different operator orderings. Observables involving higher powers of $\hat q, \hat p$ will also have several distinct orderings, and it may be that no general rule will imply the correct orderings for all observables, for a single system. If there turns out to be no single, ``correct'' ordering that works in all cases, it would mean that the program of quantization necessarily underdetermines the mathematical theory describing any particular quantum system, requiring empirical input in an essential way; knowledge of the wave function of a system, for example, would not (totally) determine the particular value of an observable with different possible orderings. Given that there is a discrete number of possible orderings, however, quantization would still present a discrete number of possible values for these observables; in this sense the theory remains predictive. On the other hand, one could argue that ordering ambiguity constitutes a \emph{prediction} of the theory, namely, that there are multiple realizations of classical quantities in the quantum domain. Nevertheless, I repeat the previously mentioned authors in their insistence that experimental input would be of great value in helping decide on the significance of the problem of operator ordering.

\section{Geometric Quantum Mechanics}

Before concluding I will describe some work done independently by Kibble (1978), Heslot (1984), Aharonov (1990), and taken up more recently by Ashtekar (1997), whose program may be called \emph{Geometric Quantum Mechanics} (GQM) \cite{Kibble:1978, Heslot:1985, Aharonov:1990, Ashtekar:1997ud}. Their work reveals a sense in which a Dirac-like rule holds \emph{exactly} in quantum theory, except that it no longer relates classical and quantum observables and therefore does not constitute a statement about the quantization map. These researchers noticed that quantum mechanics, as usually formulated, possesses a symplectic (and metric) structure of its own, and may be characterized as a Hamiltonian dynamical system, albeit a typically infinite-dimensional one. Given a Hilbert space, the symplectic structure comes from the imaginary part of the inner product, while the metric structure comes from the real part of the inner product.

There are multiple ways of uncovering the symplectic character of QM; we will follow the route of Heslot. The idea is to expand the state vector $|\psi\rangle$ in the eigenbasis of the Hamiltonian operator,
\BE
|\psi(t)\rangle = \sumop_n \lambda_n(t) | \psi_n \rangle,
\EE
and separate real and imaginary parts in the Schr\"odinger equation $i \hbar \del_t |\psi\rangle = \hat H |\psi\rangle$. Let $\lambda_n = (u_n + i v_n)/\sqrt{2\hbar}$, and let the eigenvalues of $\hat H$ be $E_n$. Then one finds the equations
\BE
\frac{\dd u_n}{\dd t} = E_n v_n, \quad \frac{\dd v_n}{\dd t} = - E_n u_n,
\EE
which have the same form as Hamilton's equations, for a ``quantum phase space'' with coordinates $u_n, v_n$ and with a quadratic Hamiltonian
\BE \label{GQMHam}
\mcal H(u,v) = \frac{1}{2} \sumop_n E_n^2 ( u_n^2 + v_n^2),
\EE
subject to the constraint that $\| \psi \|^2=1$, which is equivalently the statement that
\BE
\sumop_n ( u_n^2 + v_n^2) = 2\hbar.
\EE
That is, the phase space is, in general, a sort of infinite-dimensional sphere. Expectation values of self-adjoint operators are given by real functions quadratic in $u_n, v_n$. One can define a Poisson bracket on this space by
\BE
\{f, g\} = \sumop_n \Bigg[ \frac{\del f}{\del u_n} \frac{\del g}{\del v_n} - \frac{\del f}{\del v_n} \frac{\del g}{\del u_n} \Bigg].
\EE
The equations above may be formulated in differential-geometric terms; the Poisson bracket is then determined by the symplectic 2-form on quantum phase space, and Schr\"odinger's equation becomes the Hamiltonian flow of $\mcal H$.\footnote{The Schr\"odinger equation may also be formulated as \emph{functional} Hamilton's equations in this formalism \cite{Ashtekar:1997ud}.} Indeed, this geometric structure seems natural given that the space of distinct quantum states is not the Hilbert space, but the associated \emph{projective} Hilbert space, which is not a flat manifold. An interesting feature of GQM is then the ability to eliminate Hilbert space from the formulation of QM; one studies a particular symplectic manifold rather than dealing with vectors in Hilbert space \cite{Ashtekar:1997ud}.

It can be demonstrated that the canonical transformations on this phase space are given by the unitary transformations on the state vector $|\psi\rangle$, and that the generators of these canonical transformations are self-adjoint operators on the Hilbert space. A further identity that one can prove is the following. Let $\hat f, \hat g$ be self-adjoint operators and let $f,g$ be their expectation values in a state $|\psi\rangle$. Then it can be demonstrated that the Poisson bracket of functions on the quantum phase space equals the expectation value of the commutator of the operators corresponding to these functions. That is,
\BE
\{f, g\} = \frac{1}{i \hbar} \langle \psi | [ \hat f, \hat g ] | \psi \rangle.
\EE
Although this resembles the Dirac rule of quantization, it is \emph{not} in itself a relation between classical and quantum quantities, so it is not a statement about quantization. It does, however, conform nicely to Dirac's algebraic philosophy of QM: both QM and CM have dynamical laws that are \emph{formally} the same, though the physical meaning of the ``observables'' on their respective phase spaces differs. For example, $f(q,p)$ for a classical theory refers to some instantaneous function of position and momentum, whereas $f(u,v)$ in the quantum theory refers to some function of the state. This perspective is emphasized by Heslot in particular, when he regards CM and QM as instances of a ``generalized mechanics,'' which is specified by a phase space, a Poisson bracket, and a Hamiltonian. As mentioned above, however, GQM also identifies a metric structure associated with any quantum phase space, and it turns out that the metric may be interpreted in terms of the uncertainty principle \cite{Ashtekar:1997ud}. Thus, although CM and QM obey the same dynamical laws (formally), QM involves more structure.\footnote{Measurements may also be described geometrically in the GQM formalism \cite{Ashtekar:1997ud}.}


The program of GQM, apart from offering an interesting reformulation of QM, also points towards an avenue for generalizing QM to certain ``nonlinear'' quantum theories. The idea is to consider more complicated Hamiltonians than eq. (\ref{GQMHam}). An attempt in this direction is the work on Nambu brackets \cite{Minic:2002pd}, which replaces the oscillator Hamiltonian by a rotor Hamiltonian. See \cite{Kibble:1978} for another example. Of course, such proposals will need to be checked against experiment, but the systematic nature of this sort of generalization has a certain appeal to it, which constrains the character of possible generalizations of linear QM to remain Hamiltonian flows on quantum phase space, rather than arbitrary generalizations of the Schr\"odinger equation.

\section{Closing Thoughts}

In a letter to Schr\"odinger in May 1926, after having read his paper on the equivalence of matrix mechanics and wave mechanics, H. A. Lorentz verbalized what was so mysterious, for him, about quantization:
\begin{quote}
``In spite of everything it remains a marvel that equations in which the $q$'s and $p$'s originally signified coordinates and momenta, can be satisfied when one interprets these symbols as things that have quite another meaning, and only remotely recall those coordinates and momenta.'' (pg. 44 of \cite{Letters})
\end{quote}
We have seen that Schr\"odinger, although attempting a more principled derivation of QM through his optical-mechanical analogy, soon settled for a heuristic quantization mapping, embodied by his rule for mapping the classical Hamilton-Jacobi equation to obtain wave equations. Dirac, by contrast, elevated the similarity between CM and QM, via his algebraic philosophy, to a defining characteristic of QM, at least for systems for which his ``method of classical analogy'' was applicable. Indeed, in Dirac's approach, the algebraic philosophy provides the reason for saying that a given quantum system ``corresponds'' to some classical system to begin with. But the algebraic equivalence he initially desired (in his 1925 paper) was ultimately untenable, as we have seen via the Groenewold-Van Hove theorem, since the Dirac rule relating quantum commutators and classical Poisson brackets leads to inconsistencies, when taken with certain other desirable properties of a quantization map. Dirac himself suspected that the rule could not hold for arbitrary classical functions, as manifested later in the cautionary statement of the Dirac rule in his book. Nonetheless, a restricted use of his philosophy, where one only bothers to make the configuration coordinates and their conjugate momenta satisfy Dirac rule, has been extremely successful in practice among physicists; and it provides a starting point for finding possible quantum theories associated with any given classical theory. Whether Lorentz would have been satisfied with where things ended up is not so clear.

The drive to determine a consistent quantization mapping to ``correct'' Dirac's quantization scheme began soon after his original works, starting in 1927 with the work of Weyl and continuing ever since then. In Weyl's approach, QM is found to be algebraically equivalent not to ordinary CM, but to a \emph{deformation} of CM; thus although a consistent $\mathbb Q$ map is found, Dirac's algebraic philosophy is relaxed, and CM and QM do not obey the same laws. Even once we accept that there is no strict algebraic equivalence between CM and QM, there are reasons to be dissatisfied with WQ as the solution to the problem of quantization. First, WQ is not guaranteed to produce physically relevant operators, and second, it does not provide an explicit construction of the fundamental position and momentum operators, out of which all other operators are built, apart from providing a justification via the CCR's and appeal to irreducible action on wave functions. We have seen that the enterprise of Geometric quantization, on the other hand, does provide a constructive approach to determining these operators. Technically, however, by introducing the notion of polarization and BKS projection, GQ implements its own version of demanding irreducibility of the $\hat q, \hat p$ operators; in this sense GQ does not make more from less, compared with WQ. Still, what GQ \emph{does} provide is a direct route to the particular form of the operators in question, rather than depending on an inspired guess, by virtue of always being able to write down the prequantum operators using only the geometric structures associated with the symplectic manifold on the classical side, and building the full operators from there. GQ also makes apparent that quantum spin systems fit consistently within the program of quantization, which is not obvious at first site.

We noted in section 4 that the GVH theorem may be regarded as evidencing the absence of a  rigorous quantization map. Another way of framing this may run as follows. QM introduces a new constant of nature, $\hbar$, and thus there is ``extra information'' \cite{ZachosStackExchange} in a quantum theory than in a classical theory, or rather, QM adds a new ingredient above and beyond what was present in CM. Thus it would be strange if CM itself, via quantization, was sufficient for determining QM (apart from requiring an empirical determination of $\hbar$, once the mathematical structure is known). But we mentioned in footnote 25 the work of Hiley and de Gosson, in which WQ is used to associate classical solutions of Hamilton's equations with solutions of the Schr\"odinger equation with Weyl-quantized Hamiltonian, in a one-to-one manner. One might then conclude that a particular $\mathbb Q$ map \emph{is} sufficient to generate QM from CM.\footnote{We are tabling the issues of operator ordering and physical relevance here.} As those authors stress, however, their result does not imply that quantum theory can be derived from classical theory (or that QM \emph{reduces} to CM). As physical theories, CM and QM differ substantially in their interpretation and usage; after all, QM seems to bring in probabilities in a fundamental way which was absent in CM. Hence, we might say that the dynamical, mathematical laws of QM may be derived from CM, but not the \emph{interpretation} of the quantities appearing in those laws; both features are necessary to obtain an empirically adequate theory. 

The failure to provide the probabilistic interpretation of QM may certainly be taken to count against the program of rigorous quantization, if what one wanted was a \emph{complete} derivation of quantum theory. A natural response to this scenario is to try to derive quantum mechanics by wholly independent means from quantization, perhaps one that incorporates probability in a less ad hoc manner. We will not survey all such proposals, but content ourselves with a particular one for the sake of comparison. In Edward Nelson's Stochastic Mechanics (SM) \cite{Nelson:1966}, the Schr\"odinger equation is derived from the postulation of an underlying stochastic process that influences particle motion, analogous to a sort of universal Brownian motion, together with the postulate of a generalized Newton's Second Law. In other words, QM is derived not from ordinary classical mechanics, but from a stochastic generalization of it. In this way, probabilities enter the theory in a natural way. The occurrence of the Laplacian in the Schr\"odinger equation, under SM, results from the  diffusion term of a Fokker-Planck equation, which is generated by the underlying stochastic process, and therefore has a different relation to particle momentum than it does under the ordinary perspective of quantization; indeed, the two approaches have quite different perspectives on the relation between CM and QM. Moreover, spin has been incorporated into SM \cite{Dankel,WallstromPauli}, where it is formulated as the limit of a stochastic process on $\mathbb R^3 \times SO(3)$ configuration space as the moment of inertia of the spinning object goes to zero. However, I believe it is fair to say that the program of quantization is more adaptable, being more easily employable when confronted with the problem of formulating a new quantum system, than trying to begin by formulating from the Nelsonian perspective. This does not necessarily imply that a Nelsonian account cannot always be given, or that such an account is not valuable.

In closing, I remark that another response, though perhaps a peculiar one, to the aforementioned interpretational deficiencies of quantization, is to deny the fundamental role of probabilities in quantum theory from the start, deeming them only necessary for \emph{us} to interpret the theory. This may be reflected in Dirac's own work on transformation theory: interestingly, in the conclusion to his paper from December 1926, ``The Physical Interpretation of the Quantum Dynamics,'' he hinted that he believed QM was not fundamentally about probabilities. This, of course, was after Born's paper where the statistical interpretation of wave functions was introduced. After having described the equivalence of his transformation theory to the probabilistic interpretation, he writes,
\begin{quote}
``In conclusion it may be mentioned that the present theory suggests a point of view for regarding quantum phenomena rather different from the usual ones [\dots] The notion of probabilities does not enter into the ultimate description of mechanical processes; only when one is given some information that involves a probability [\dots] can one deduce results that involve probabilities.'' (pg. 229 of \cite{DiracBook})
\end{quote}
It is not clear if the absence of a probabilistic aspect to the program of quantization factored into his motivation for writing those lines, though it does seem consistent with the sentiment he expresses. It is also consistent with his previous insistence that the algebraic aspect of the laws of QM were their fundamental characteristic, and is reminiscent of his initial reluctance to specify the nature of the q-numbers beyond their algebraic properties. However, because the predominant view of quantum theory incorporates probabilities in an essential way, it is not likely that this response to the deficiency problem of quantization would be popular.

\paragraph{Acknowledgments.} This work is an outgrowth from my earlier studies in Geometric Quantization, to which I was first pointed by Markus Pflaum. It has benefited from comments received after giving talks to the Rutgers Online Research Group in Mathematical Physics as well as the 20th European Conference on Foundations of Physics, 2021. I would like to thank David Wallace and Peter Woit for insightful conversations during that conference. I would also like to thank Maaneli Derakhshani for several useful discussions and for alerting me to the Nelsonian approach to spin.


\bibliographystyle{ieeetr}
\bibliography{QM_Refs.bib}

\end{document}